\documentclass[11pt,aps,prd,letterpaper,preprintnumbers,floatfix,nofootinbib,showpacs,superscriptaddress, notitlepage]{revtex4-1} \usepackage{amssymb}
\usepackage{slashed}
\usepackage{amsmath}
\usepackage{mathtools}
\usepackage{float}
\usepackage[utf8]{inputenc}
\usepackage{graphicx}
\usepackage{latexsym}
\usepackage{natbib}
\usepackage[normalem]{ulem}

\renewcommand{\mtt}{\ensuremath{m_{\tau\tau}}}

\newcommand{\eLpR}{\ensuremath{\mathrm{e^-_{L80} e^+_{R30}}}}

\newcommand{\eRpL}{\ensuremath{\mathrm{e^-_{R80} e^+_{L30}}}}

\newcommand{\eett}{\ensuremath{e^- e^+ \to \tau^- \tau^+}}
\newcommand{\tpn}{\ensuremath{\tau^\pm \to \pi^\pm \nu}}
\newcommand{\trn}{\ensuremath{\tau^\pm \to \pi^\pm \pi^0 \nu}}

\date{\today}

\begin{document}

\title{Measuring the tau polarization at ILC}
\author{Keita Yumino}
\affiliation{The Graduate University for Advanced Studies, SOKENDAI}
\author{Daniel Jeans}
\affiliation{KEK, Tsukuba, Japan}
\affiliation{The Graduate University for Advanced Studies, SOKENDAI}

\begin{abstract}

Measurement of the tau lepton polarization in \eett\ is an important electro-weak measurement at ILC and other
future electron-positron colliders. In this paper we discuss several methods to extract polarimeter
information for \eett\ events at the nominal centre-of-mass energy, and develop a new method, based
on charged particle impact parameter measurement, which can accurately reconstruct tau momenta
even in events with significant Initial State Radiation.

In future work we will extend the study to estimate the precision with which the tau polarization can be measured at ILC-250,
both for high-mass tau pairs and for those which radiatively return to the $Z^0$ peak.
This will complement our past study which showed that the tau polarization can be measured to better than 1\% at the ILC-500.

\end{abstract}

\maketitle

\vskip -0.3in
\begin{center}
\rule[-0.2in]{\hsize}{0.01in}\\
 \vskip 0.1in
 Submitted to the  Proceedings of the US Community Study\\
 on the Future of Particle Physics (Snowmass 2021)\\
 \rule{\hsize}{0.01in}\\
\end{center}

\section{Introduction}

The polarized beams foreseen for the ILC allow sensitive probes of Left and Right-handed electrons' neutral current couplings, both at the 
nominal collision energies (250, 350, 500, 1000~GeV) and in the ``radiative return'' to the $Z$ boson.
At a collider with unpolarized beams, for example at LEP, measurement of the tau lepton polarisation in tau-pair events provides a similar
measurement of the chirality of the neutral current interactions.
At the ILC, it will be possible to simultaneously use both of these methods.
The relation between these two measurements will be affected both by the true beam polarization -- providing a powerful check for systematic effects --
and the universality of the chiral interactions between lepton generations -- a particularly interesting topic in these days of flavor anomalies.

We discuss studies of the extraction of tau polarisation information at an electron positron collider such as the ILC.
Several different methods to extract polarisation observables are studied.

\section{Polarimeters}

\begin{figure}[htb]
   \centering
   \includegraphics[width=0.4\textwidth]{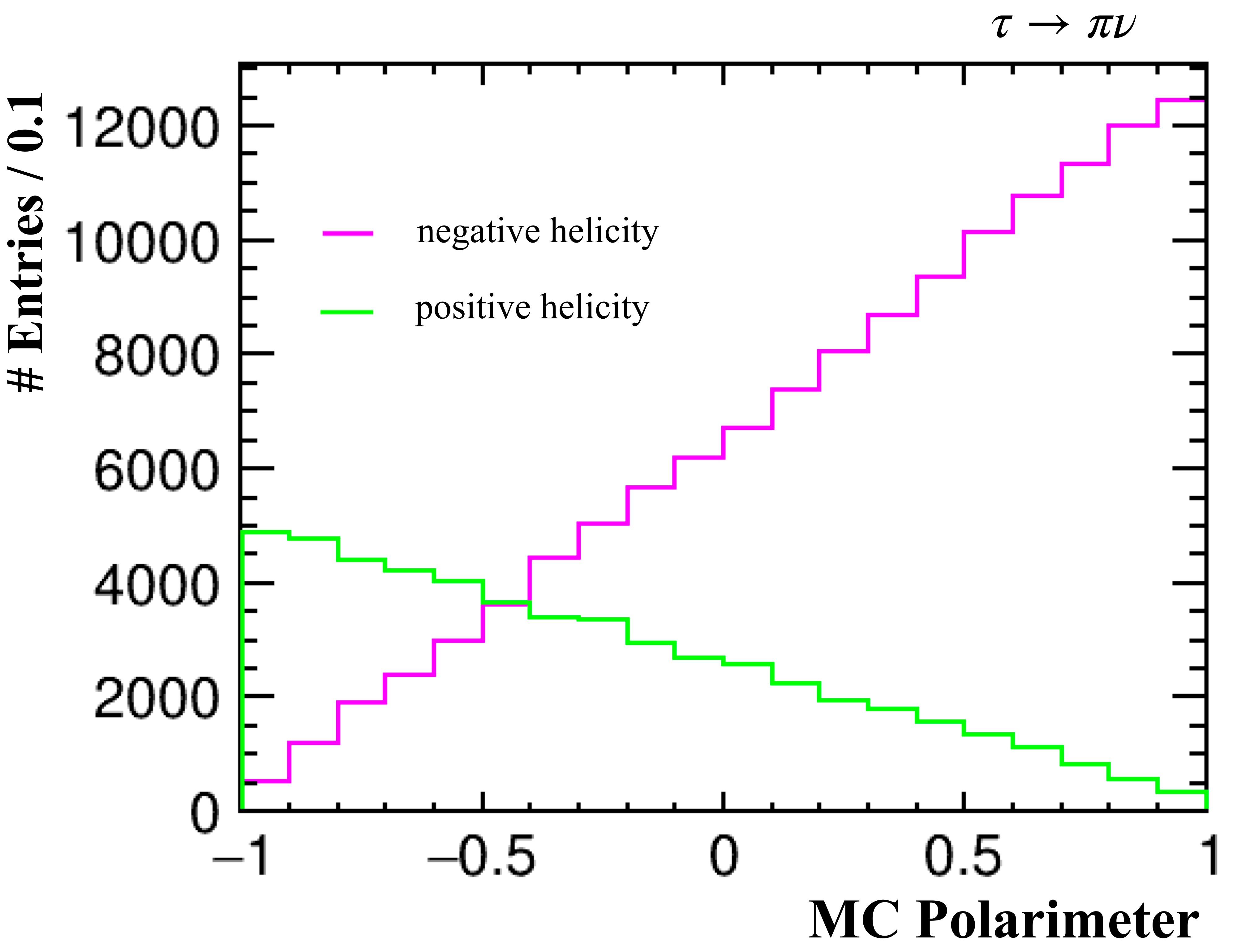}
   \includegraphics[width=0.4\textwidth]{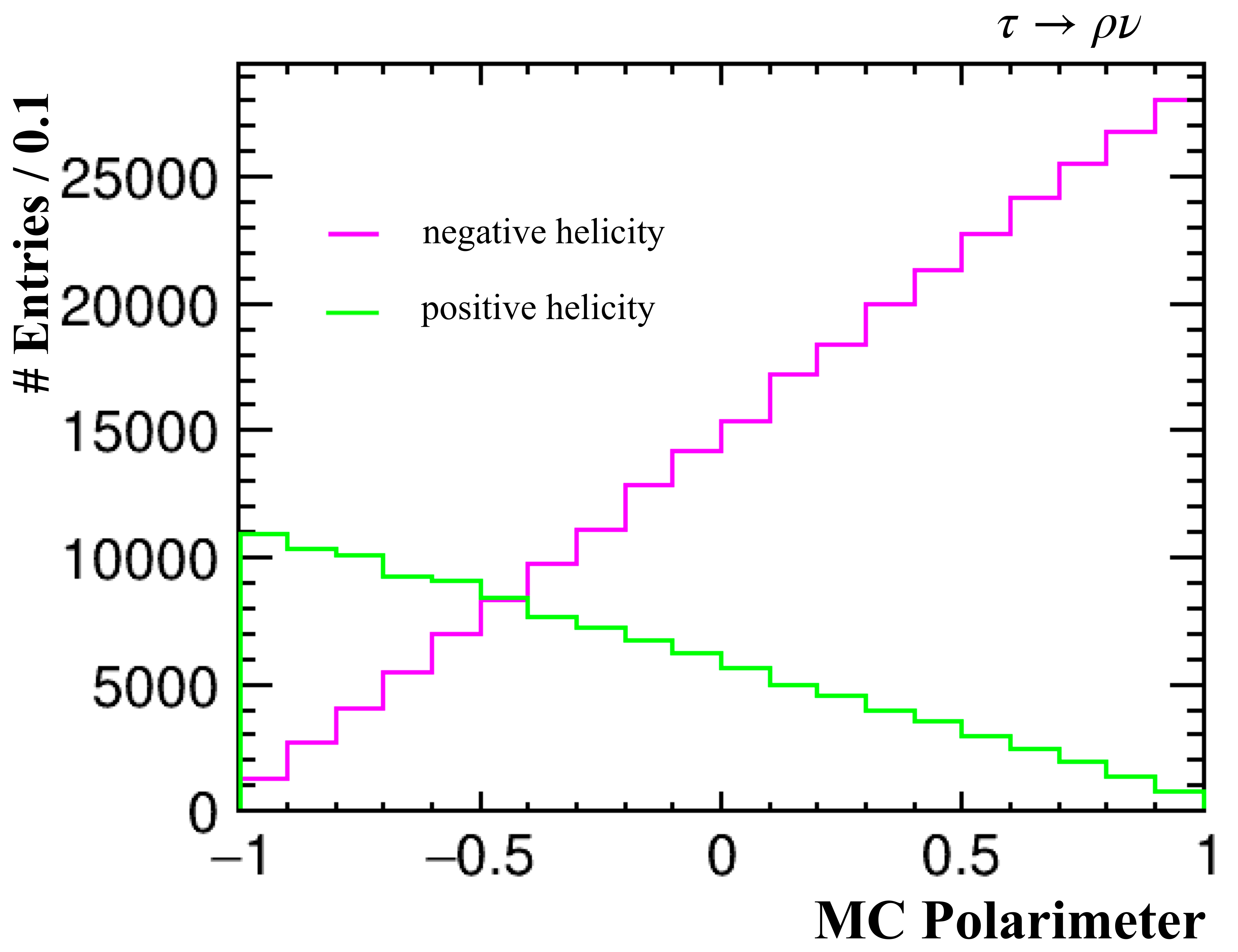} \\
   \caption{
     ``Optimal'' polarimeter distributions for \tpn\ (left) and \trn\ (right) decays of $\pm$ helicity tau leptons,
reconstructed using MC truth information.}
   \label{fig:polmc}
\end{figure}

The distribution of a tau lepton's decay products depends on the orientation of its spin.
Polarization-sensitive observables, known as polarimeters, make use of this dependence to reconstruct the 
spin orientation.
We study hadronic decays of the tau, which offer greater sensitivity to the tau's polarisation.
Optimal tau polarimeter vectors can be rather simply defined in the case of \tpn\ (which we sometimes abbreviate as ``$\tau \rightarrow \pi$'') and 
\trn\  (``$\tau \rightarrow \rho$'') decays, see e.g.~\cite{Tauola}.
The polarimeter vectors are defined in the tau rest frames as follows: for \tpn, it is the direction of the neutrino momentum,
while for \trn\ it is the direction of the vector $\mathbf{P} = 2 (\mathbf{q} \cdot \mathbf{p_\nu}) \mathbf{q} - m_q^2 \mathbf{p_\nu}$, where 
$\mathbf{q} = \mathbf{p}_{\pi^\pm} - \mathbf{p}_{\pi^0}$, and $\mathbf{p_\nu}, \mathbf{p}_{\pi^\pm}, \mathbf{p}_{\pi^0}$ are respectively
the 3-momenta of the neutrino, charged and neutral pions. 
To distinguish taus of different helicity, we consider the cosine of the angle this polarimeter vector makes to the tau flight direction: we call this the ``polarimeter''.
We refer to this form of the polarimeters as ``optimal''; example distributions are shown in Fig.~\ref{fig:polmc}.
Extracting this optimal form of the polarimeter requires knowledge of the tau neutrino momentum, 
which is obviously not directly measurable, or equivalently of the parent tau momentum. 
It also requires the reconstruction of every visible particle produced in the
tau decay, and the correct identification of the tau decay mode.

``Approximate'' polarimeters can be defined, reconstructed based only on the momenta of visible tau decay products. 
In the 2-body decay \tpn, 
the energy of the pion is an optimal polarimeter for taus of known energy, however due to 
the spread in beam energies at a real collider and the resulting spread in tau energies, its sensitivity is slightly decreased.
In the case of \trn, one can arrive at an approximate polarimeter by integrating over possible
neutrino momenta, as described in \cite{duflot}. 
In the case of \tpn\ decays, the approximate method retains almost all the sensitivity of the ``optimal'' analysis, 
while for \trn\ decays the sensitivity of the ``approximate'' method is significantly smaller than the ``optimal'' one.

\section{Approximate polarimeters at ILC-500}

High-mass tau pairs at ILC-500 provide an experimental challenge due to the large boost, leading to very highly collimated tau jets whose 
constituents can be challenging to resolve. 
We here summarize results presented in~\cite{Jeans:2019brt}.

Tau-pair event samples were generated using \texttt{WHIZARD} version 1.95~\cite{whizard, omega}, 
taking into account the ILC beam energy spread, beamstrahlung and initial state radiation. 
After decaying the taus using \texttt{TAUOLA}~\cite{Tauola}, events were fully simulated and reconstructed using standard ILD tools based on 
\texttt{ddsim/DD4hep}~\cite{dd4hep} and \texttt{MarlinReco}~\cite{marlinreco}.
The principal output of the reconstruction is a collection of Particle Flow Objects (PFO), corresponding to reconstructed
final state particles.

An event selection was developed to select high-mass pairs of taus, and to reject other processes.
A selection efficiency of around 60\% was achieved for high-mass tau pairs with at least one hadronic tau decay,
with around 1.5--4\% contamination from non-tau backgrounds, depending on the assumed beam polarisation.
The separation of the single-prong hadronic decay modes to one, two, or three pion modes achieved an efficiency
of between 90\% and 60\%, depending on the decay mode.
One-- and two--pion decays were used to calculate ``approximate'' polarimeters from reconstructed information. 
Pseudo-experiments carried out on simulated
polarimeter distributions were used to estimate the statistical precision achievable at ILC-500: better than
1\% for the 1.6/ab of data in each of the \eLpR\ and \eRpL\ polarisations, and around 2\% in the 
like-sign polarisations for which 0.4/ab was assumed.

\begin{figure}[htb]
  \includegraphics[width=0.4\textwidth]{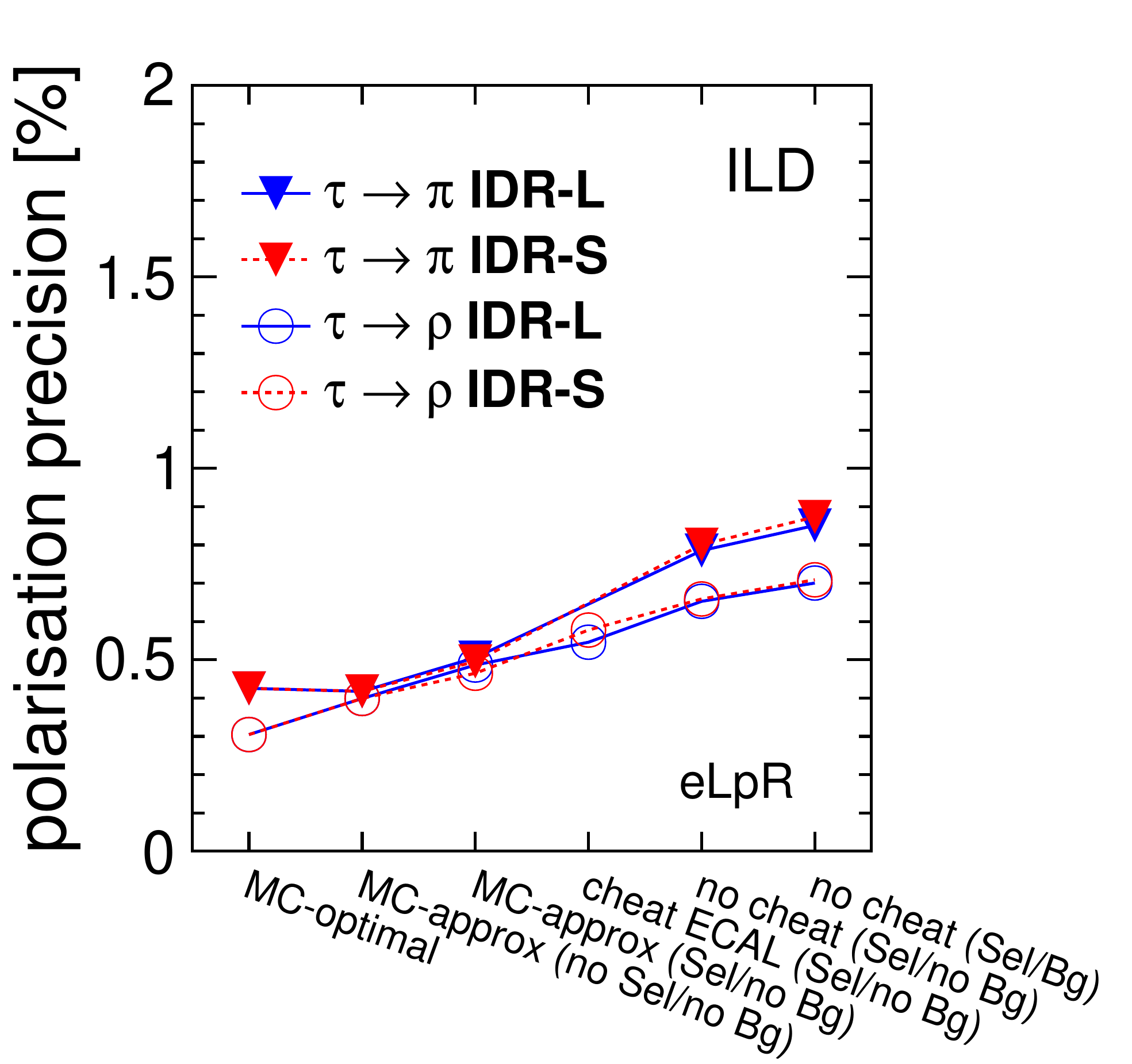}
  \includegraphics[width=0.4\textwidth]{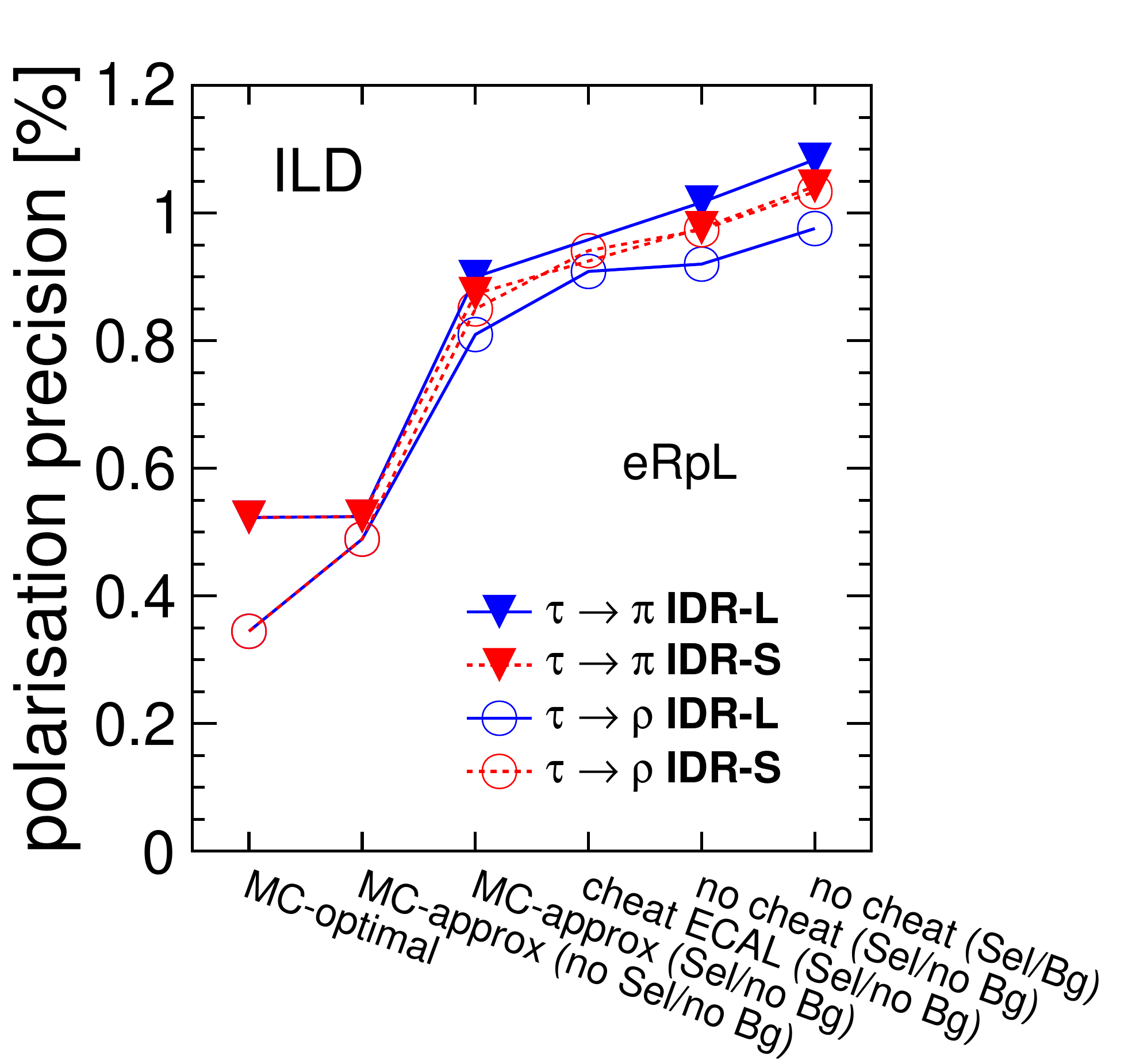} \\
  \caption{Statistical precision on the tau polarisation at ILC-500, in the \eLpR\ and \eRpL\ polarizations,
at various levels of ``cheating'', from optimal MC information (``MC-optimal'') to full simulation and realistic reconstruction (``no cheat (Sel/Bg)'').
Triangles (open circles) correspond to the \tpn\ (\trn) decays, and IDR-L/S refer to two ILD models~\cite{Jeans:2019brt}.
}
  \label{fig:chealPolPrec}
\end{figure}

The influence of various experimental effects was also investigated: use of the ``approximate'' form of the polarimeters; 
signal inefficiency and background contamination; and ECAL energy resolution, shown in Fig.~\ref{fig:chealPolPrec}. Each of these factors gives rise to 
a similar level of degradation in experimental precision, which is degraded by a factor of around two compared to the
ideal result.

\section{Reconstruction of the tau momentum in di-tau events}

To go beyond the approximate polarimeters described above, explicit reconstruction of the tau momentum is required.
We discuss several ways in which the tau momenta can be directly reconstructed in the \eett\ process.
We study these methods using the MC truth information of di-tau events generated by WHIZARD, taking
into account the effects of ISR and the beam energy spread and beamstrahlung expected at ILC-250.
Tau leptons are decayed by TAUOLA. No detector simulation or realistic reconstruction algorithms were applied.

\subsection{Cone method}

In the case of a di-tau event both of whose taus decay hadronically, 
if the two tau leptons are back-to-back and each carry the nominal beam energy, the tau momenta can
be calculated to within a 2-fold ambiguity. 
These assumptions are applicable to tau pairs produced at the nominal centre-of-mass energy, with no energy
loss via ISR or beamstrahlung.
The assumed energy and invariant mass of the tau define both the
energy of the neutrino and the angle which it makes to the visible tau momentum. Equivalently,
one can calculate the angle which the tau momentum makes to the visible momentum, defining a cone around
the visible momentum on which the tau momentum lies.

\begin{figure}[htb]
  \centering
  \includegraphics[width=0.6\textwidth]{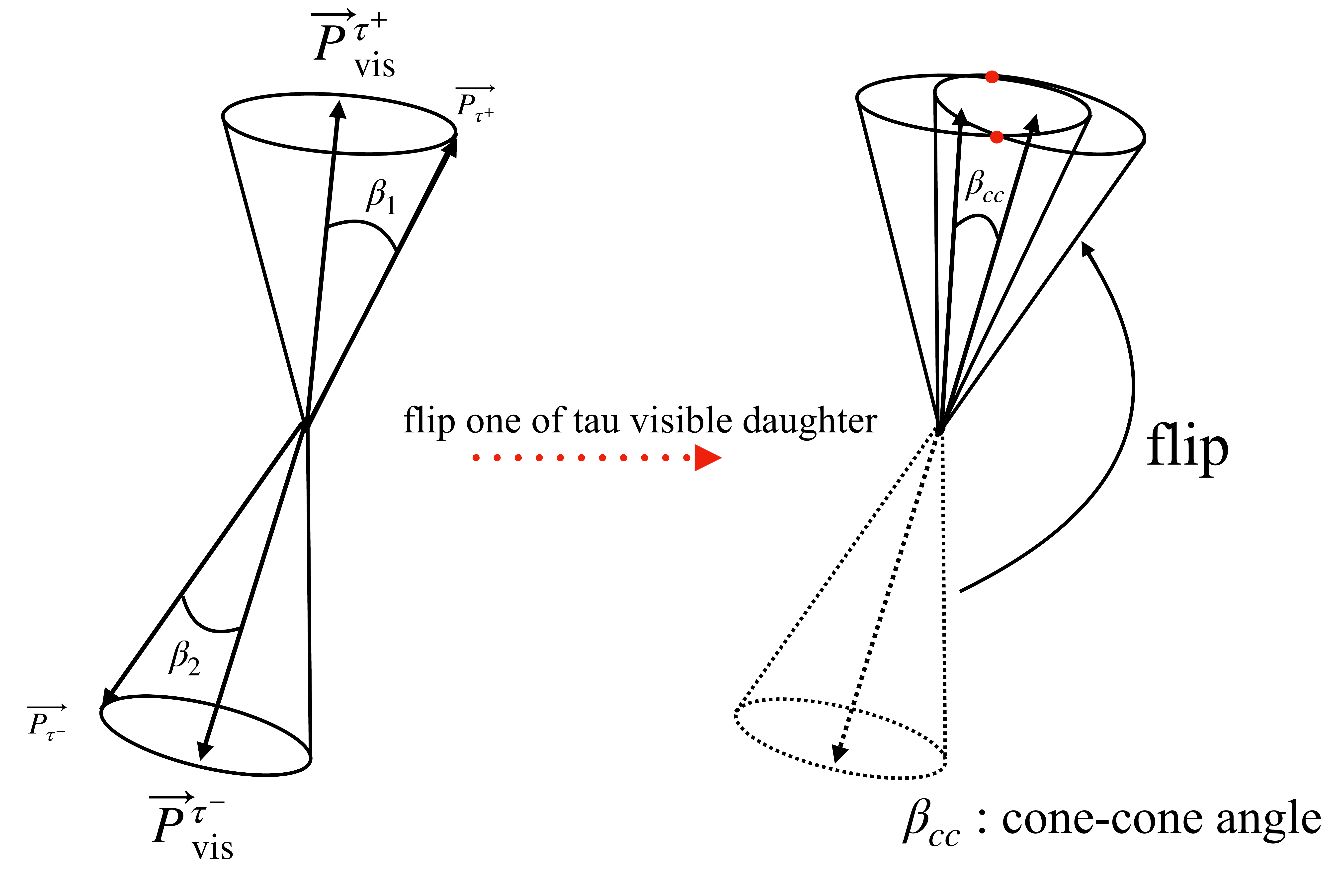}
  \caption{
    Illustration of the cone method, showing the cone intersections corresponding to the tau momentum solutions.
  }
  \label{fig:conemeth}
\end{figure}

The two taus in an event are assumed to be back-to-back in the collision frame. The intersections of the two cones in
an event then represent tau momentum directions which are consistent with all assumptions and constraints. 
The number of possible solutions is either 2 (in the case of intersecting cones, as illustrated in Fig.~\ref{fig:conemeth}) or 0 (if the cones do not intersect).
The assumptions made above are not valid in the case of significant initial state radiation, or non-nominal
beam energy, for example due to beamstrahlung. Under such conditions, the cones often do not intersect,
leading to an inefficiency of the method, or, if they do intersect, to inaccurate reconstruction.

%
%
%

\begin{figure}[htb]
  \centering
  \includegraphics[width=0.4\textwidth]{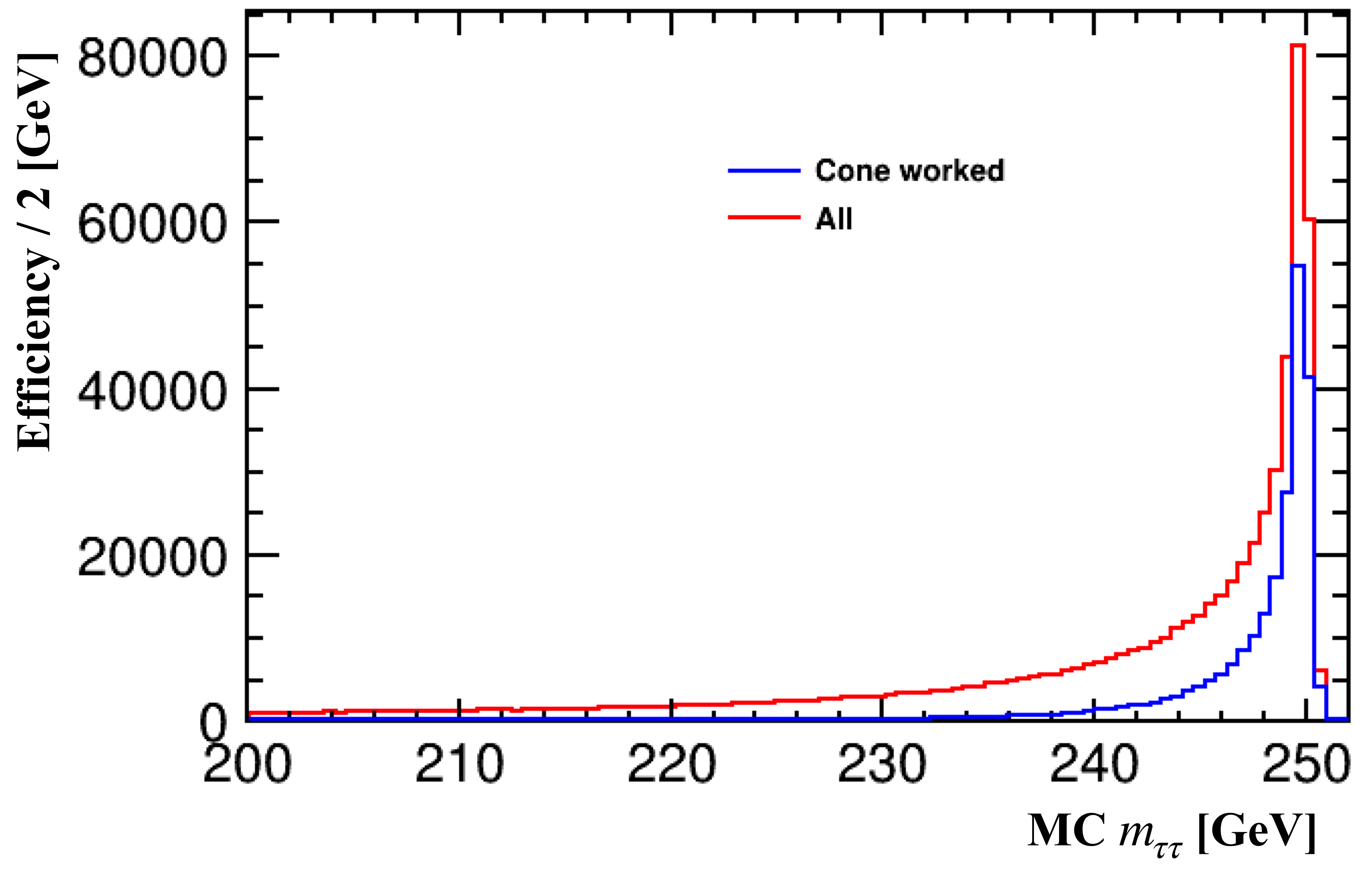}
  \includegraphics[width=0.4\textwidth]{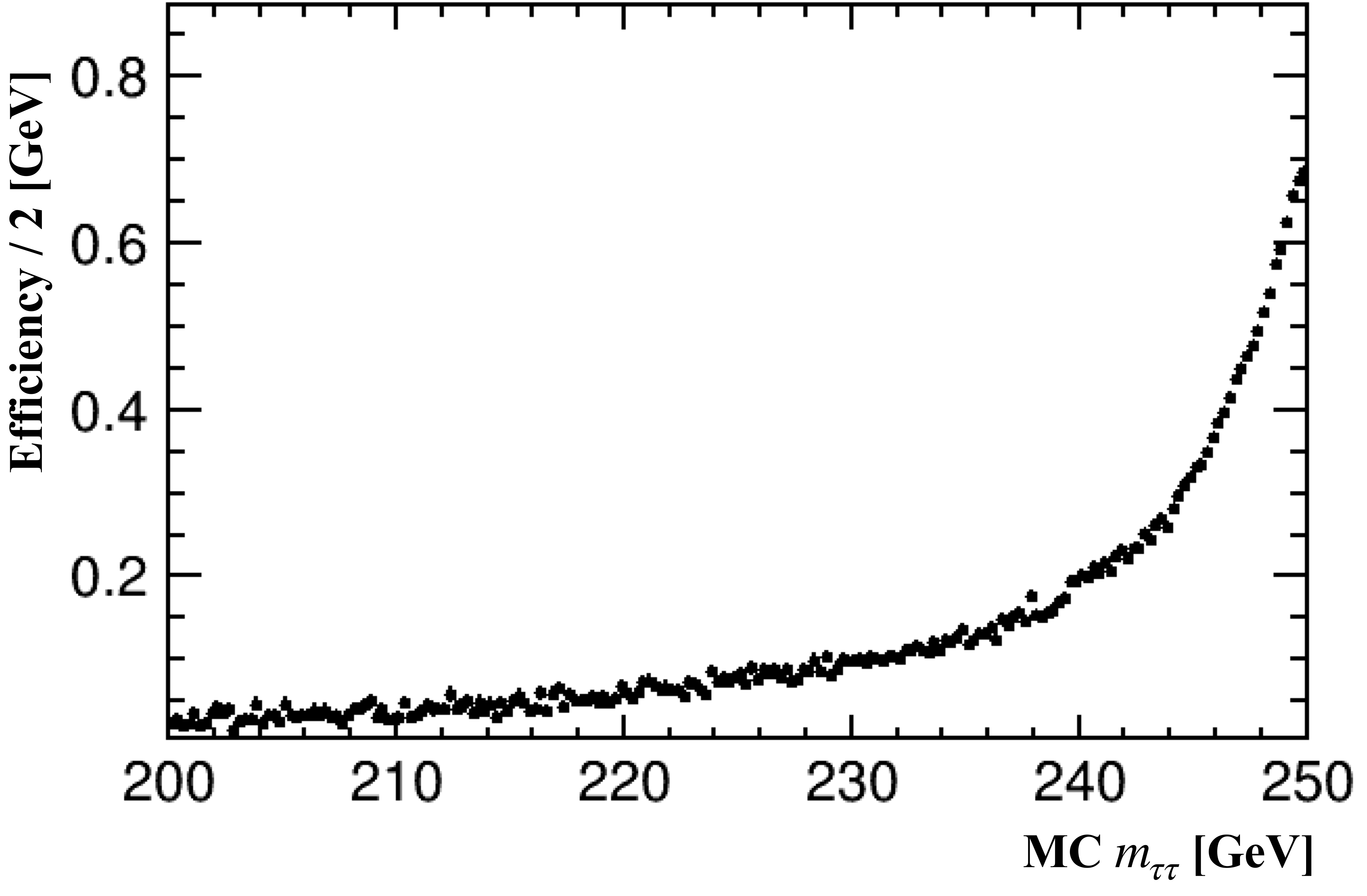} \\
  \caption{Distribution of \eett\ events in their invariant mass \mtt, for all events and for those in which
the cone method identifies a solution. The right plot shows the cone method's efficiency as a function of \mtt.}
  \label{fig:ConeEff}
\end{figure}

Fig.~\ref{fig:ConeEff} shows the original distribution of events in the tau invariant mass \mtt\ and the distribution of events 
for which the cone method found a solution. The ratio of the two is also shown, demonstrating that while the 
cone method has a reasonable efficiency of $\sim70\%$ for events at $\mtt \sim 250~$GeV, the efficiency drops to around 
$20\%$ by $\mtt=240~$GeV and $< 5\%$ by $\mtt=200~$GeV.
Even a rather modest deviation from the method's assumptions can have a strong effect on the ability to identify a solution.

For events for which the cone method does not find a solution, one possible option is to take the midpoint of cone surfaces as a solution for the tau
momentum direction. We call this ``Midpoint method''. This can always find a solution, however it may be far from the true one, particularly for events
whose taus are in reality not back-to-back, such as those preceded by ISR.



\begin{figure}[htb]
   \centering
   \includegraphics[width=0.4\textwidth]{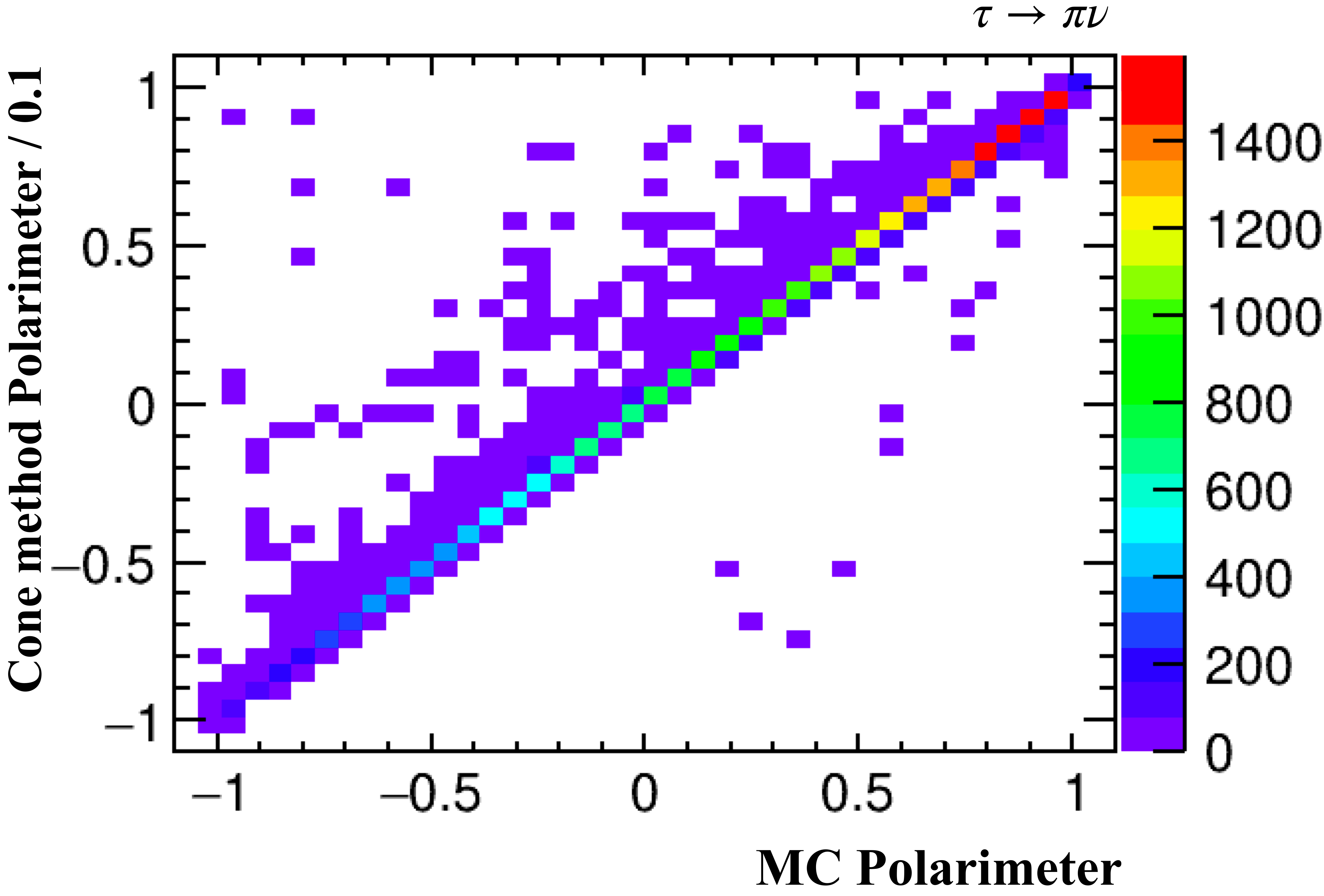}
   \includegraphics[width=0.4\textwidth]{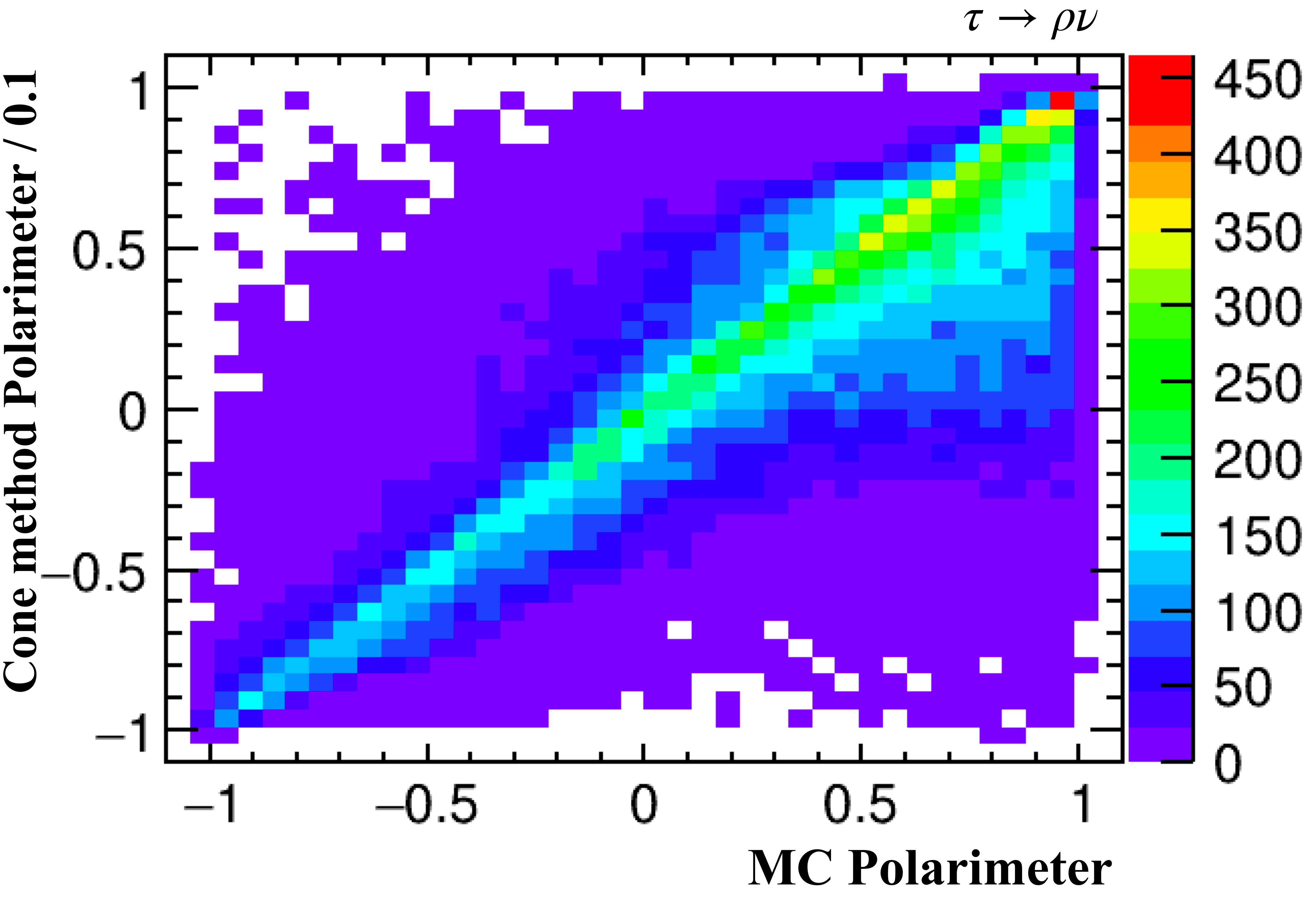} \\
   \includegraphics[width=0.4\textwidth]{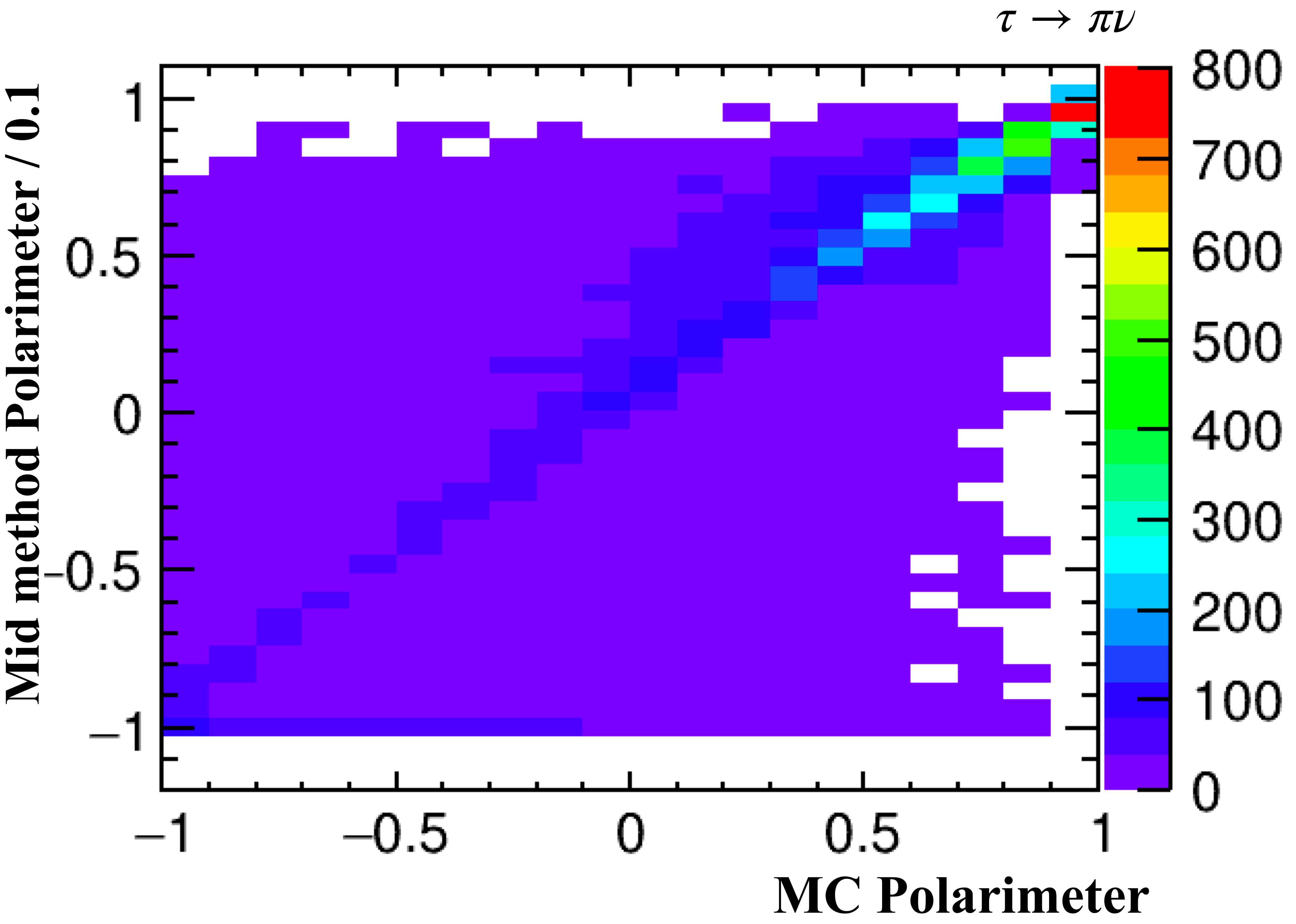}
   \includegraphics[width=0.4\textwidth]{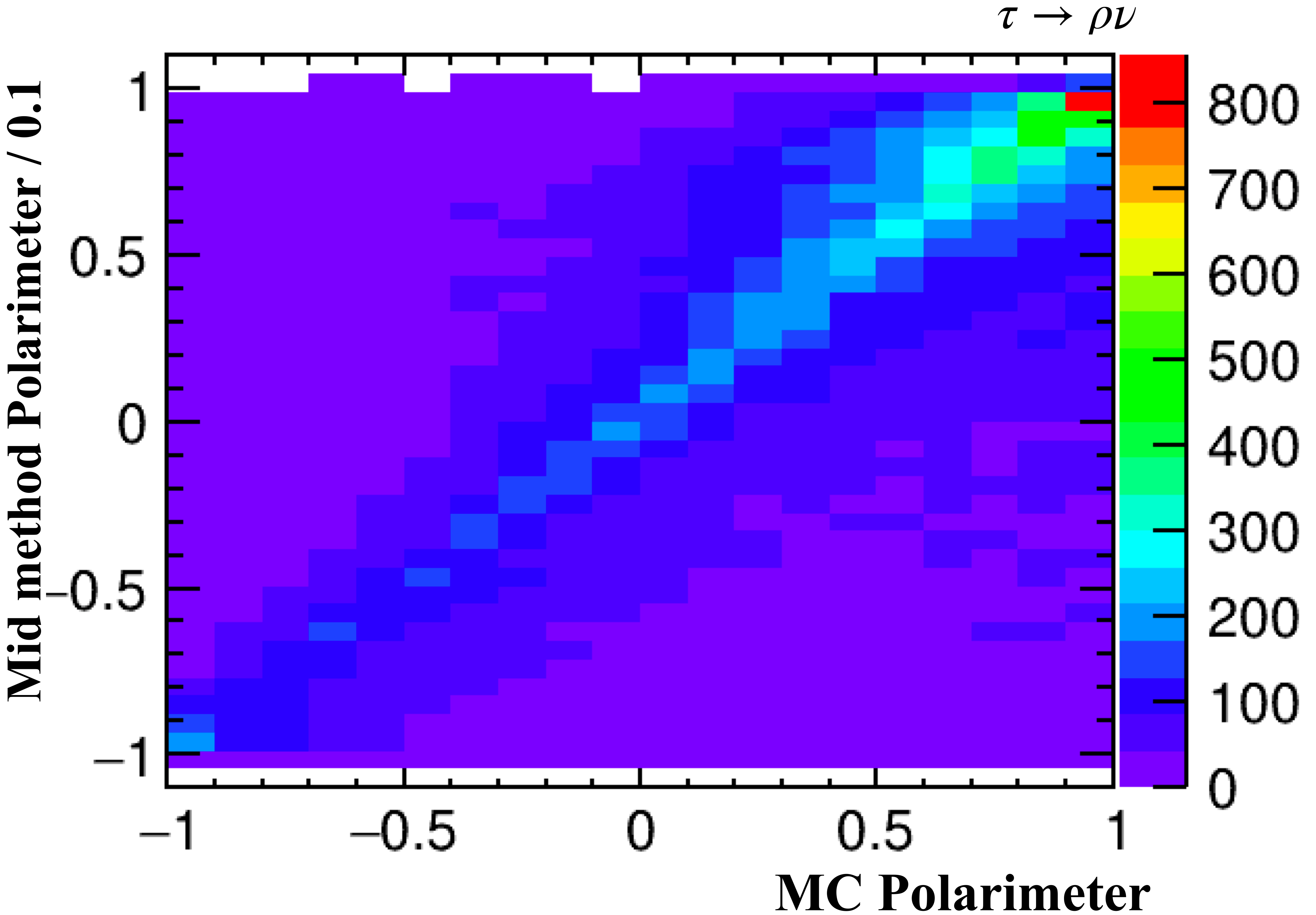} \\
   \caption{
Correlation between reconstructed and true polarimeters
for \tpn\ (left) and \trn\ (right) decays,
for events reconstructed using the cone (upper), and midpoint (lower) methods. 
}
   \label{fig:polcomp}
\end{figure}

Figure~\ref{fig:polcomp} compares the polarimeters calculated using the cone and midpoint methods with the MC truth values.
In the case of \tpn\ decays, the polarimeter reconstructed using the cone method is in good agreement with the optimal one, while for \trn\ 
decays the reconstruction is noticeably worse. Events in which the cone method fails -- predominantly those with \mtt\ far from the nominal 
centre-of-mass energy -- are then passed to the midpoint method. Comparisons of the extracted polarimeter with the MC truth for these
events are shown in the same figure. The accuracy of polarimeter reconstruction is markedly worse, since the assumptions are, in general,
broken more severely for such events.


\subsection{Impact parameter method}

We here describe a new tau reconstruction method which is much less sensitive to ISR, which allows reconstruction of tau momenta and
optimal polarimeters for events at different effective centre-of-mass energies.
This is achieved by including information of tau decay daughters' trajectories close to the interaction point: in other words, their impact parameters.
The vertex detectors being designed for ILC have superb single hit resolution, low mass to minimize multiple scattering, and will be placed
within $\sim 15$~mm of the interaction point. This provides an excellent measurement of the impact parameter. Use of the impact parameter to
reconstruct tau leptons from Higgs boson decay has already been demonstrated in~\cite{Jeans:2015vaa}. 
The event topology in \eett\ is significantly different, so the methods therein developed are not directly applicable.

The ILC and other electron-positron Higgs factories typically have an interaction region that has a very small transverse size
(significantly smaller than the impact parameter resolution), and somewhat long in the $z$ direction along the beams.
In the case of the ILC, the beam size at 250 GeV will be $7~nm \times 700~nm \times 0.3~mm$.
We therefore assume that the primary interaction occurs along the beam line $x=y=0$, and that $z$ component of the interaction point is unknown.

We additionally assume that ISR has negligible transverse momentum
(or in other words that the taus are back-to-back in the x-y projection of the nominal CM frame),
and that any energy loss due to ISR can be modelled as a single photon collinear with one of the incoming beams.

\begin{figure}[htb]
  \centering 
  \includegraphics[width=0.6\textwidth]{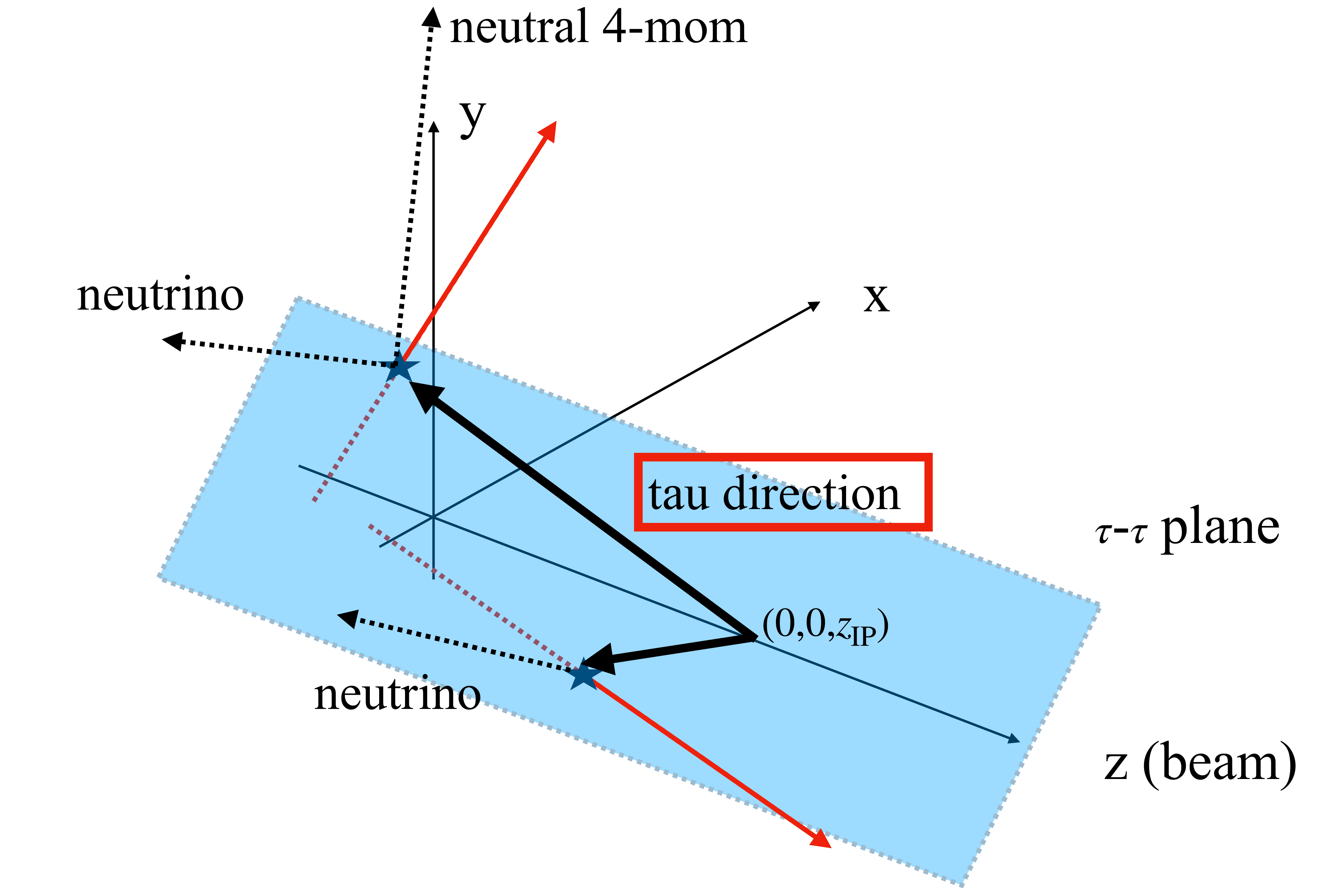}
  \caption{Schematic image of calculating tau direction.}
  \label{cartoon}
\end{figure}

We then consider a plane which contains the $z$ axis (ie the beam axis), at some angle $\phi$ to the $x$ axis, as illustrated in Fig.~\ref{cartoon}.
The interaction point lies somewhere along the $z$ axis, and therefore within the plane. Since the taus are back-to-back, their momenta also 
lie in the plane at a particular value of $\phi$.
The intersection of the trajectories of charged tau daughters with this plane can be calculated
(in this study we assume these trajectories can be approximated as straight lines near the IP);
these intersections correspond to the decay points of the tau leptons.

If we take some general interaction position along the $z$-axis, defined by its coordinate $z_{\rm{IP}}$, 
then the flight direction of the taus is defined, 
being between the intersection point and the decay points.

\begin{figure}[H]
  \centering 
  \includegraphics[width=0.9\textwidth]{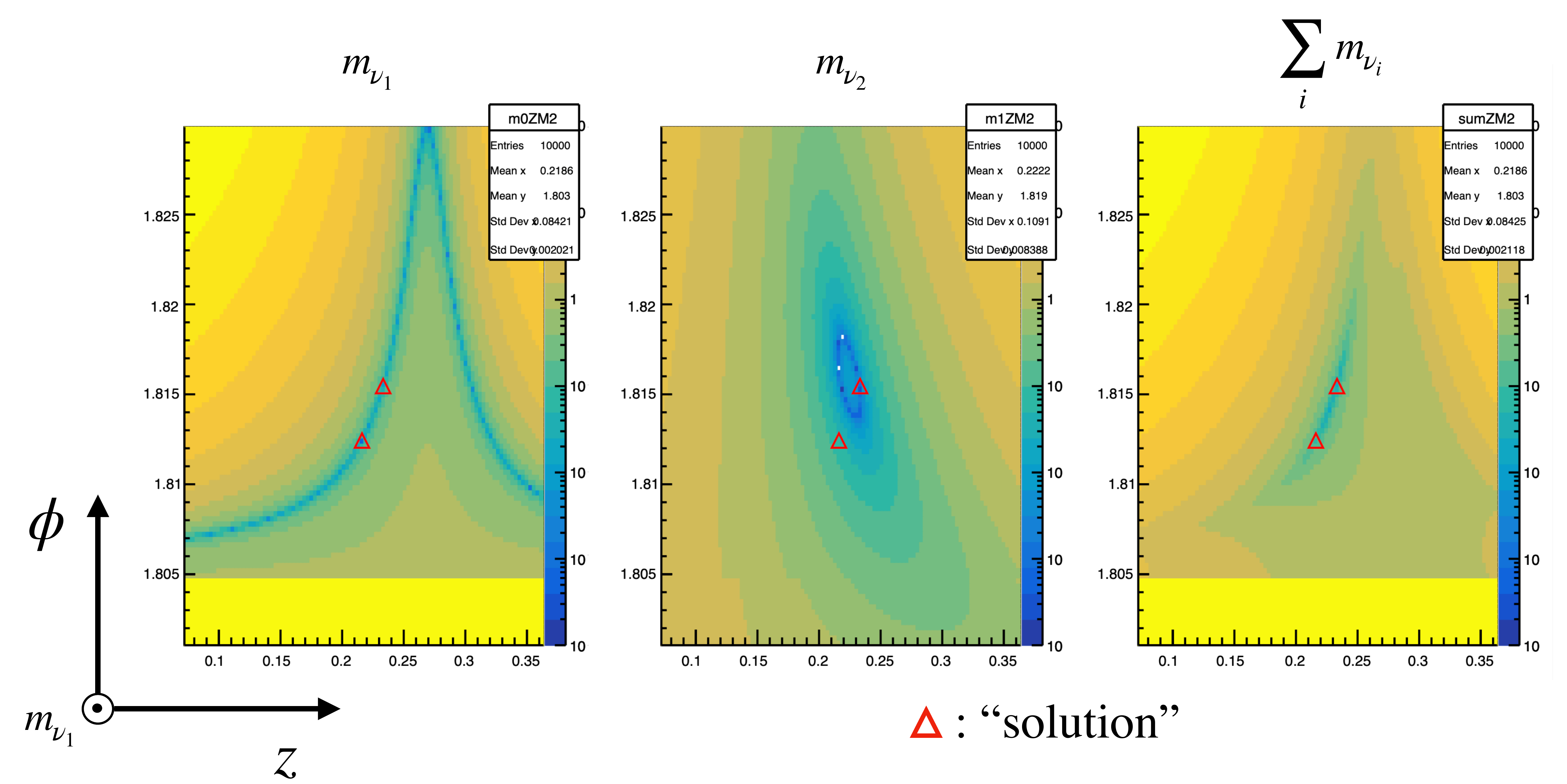}
  \caption{Dependence of the invisible masses on $\phi$ and $z_{\rm{IP}}$. 
    The color scale represents the invisible invariant mass of the two taus (left, center) and their sum (right).
    Red triangles correspond to the identified solutions, local minima in the right-hand plot.
  }
  \label{choose}
\end{figure}

For each choice of $\phi$ and $z_{\rm{IP}}$ the two tau 4-momenta can be calculated by applying 4-momentum conservation and
assuming a single ISR photon.
By comparing this calculated 4-momentum to the visible tau 4-momentum, 
the invariant mass of each tau's missing momentum (i.e. the mass of the invisible tau decay products) can be calculated.
Since we are concerned mostly with hadronic tau decays,
we choose the values of $\phi$ and $z_{\rm{IP}}$ which result in neutrino masses closest to zero.
The equations are difficult (even impossible?) to solve analytically, so we use a graphical method to find solutions.
Figure~\ref{choose} shows the extracted invisible invariant masses for the two taus depending on the choice of $\phi$ and $z_{\rm{IP}}$,
and the positions of the identified local minima which are taken as potential solutions.

\begin{figure}[htbp]
  \centering
  \includegraphics[width=0.4\textwidth]{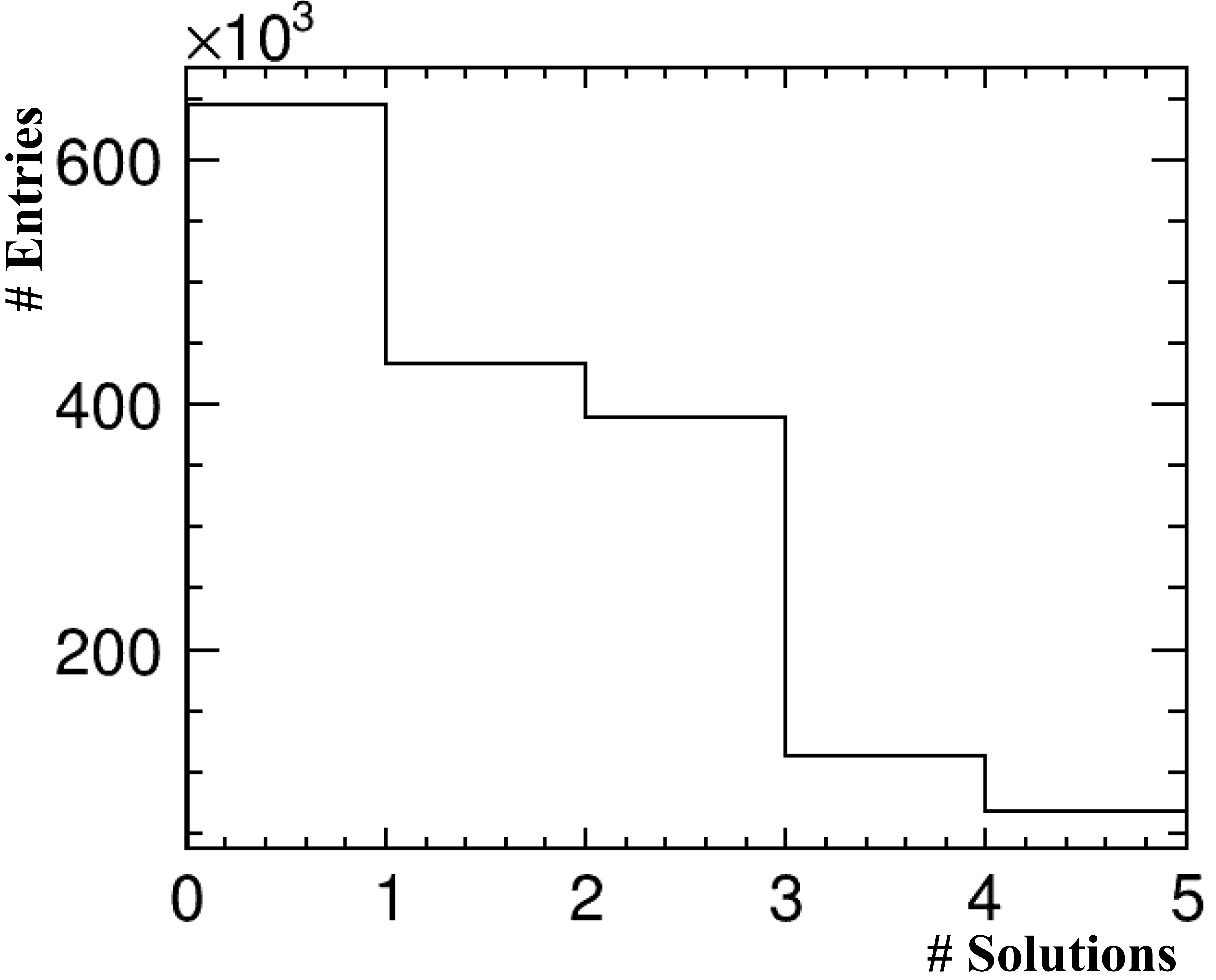}
  \includegraphics[width=0.4\textwidth]{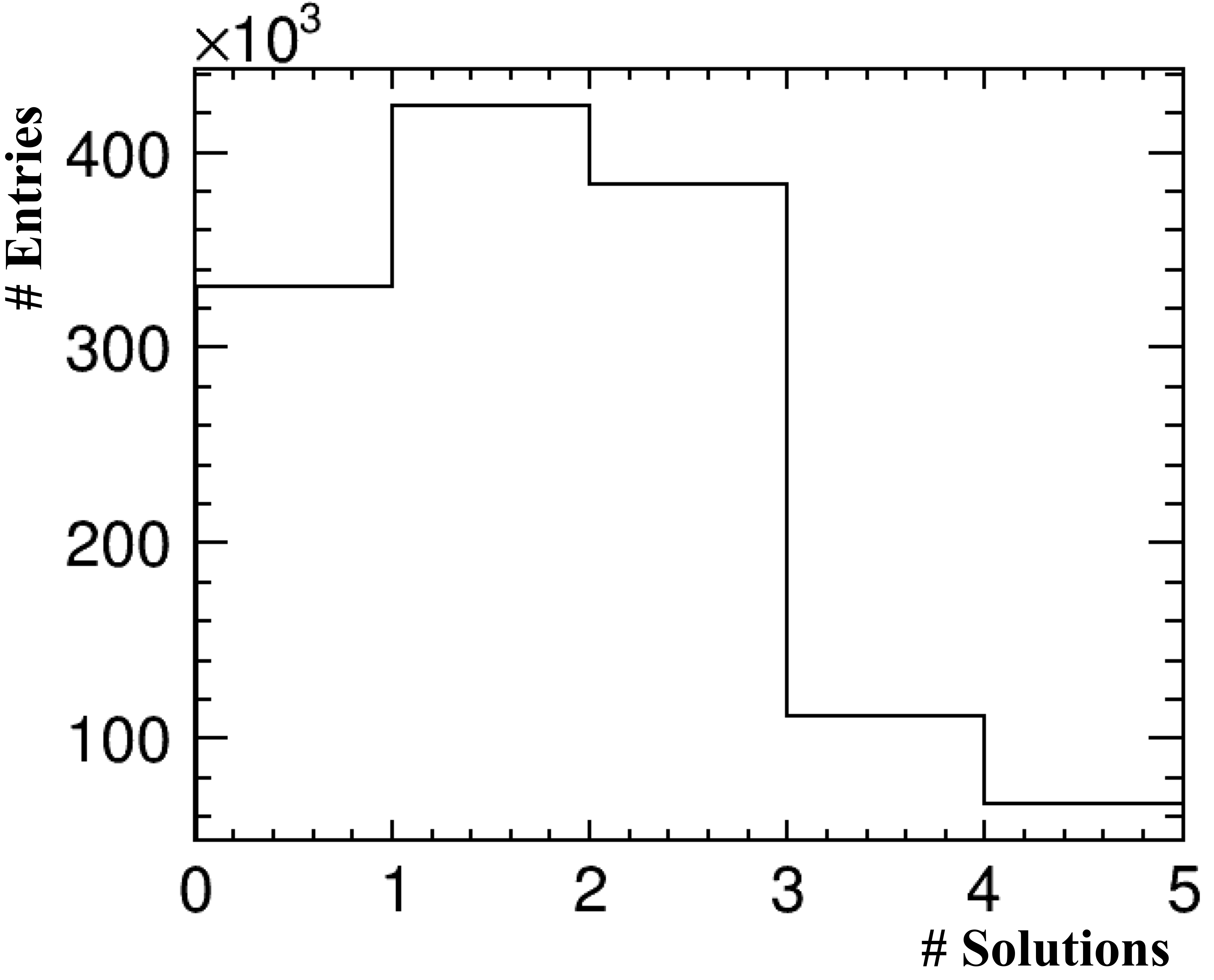} \\
  \includegraphics[width=0.4\textwidth]{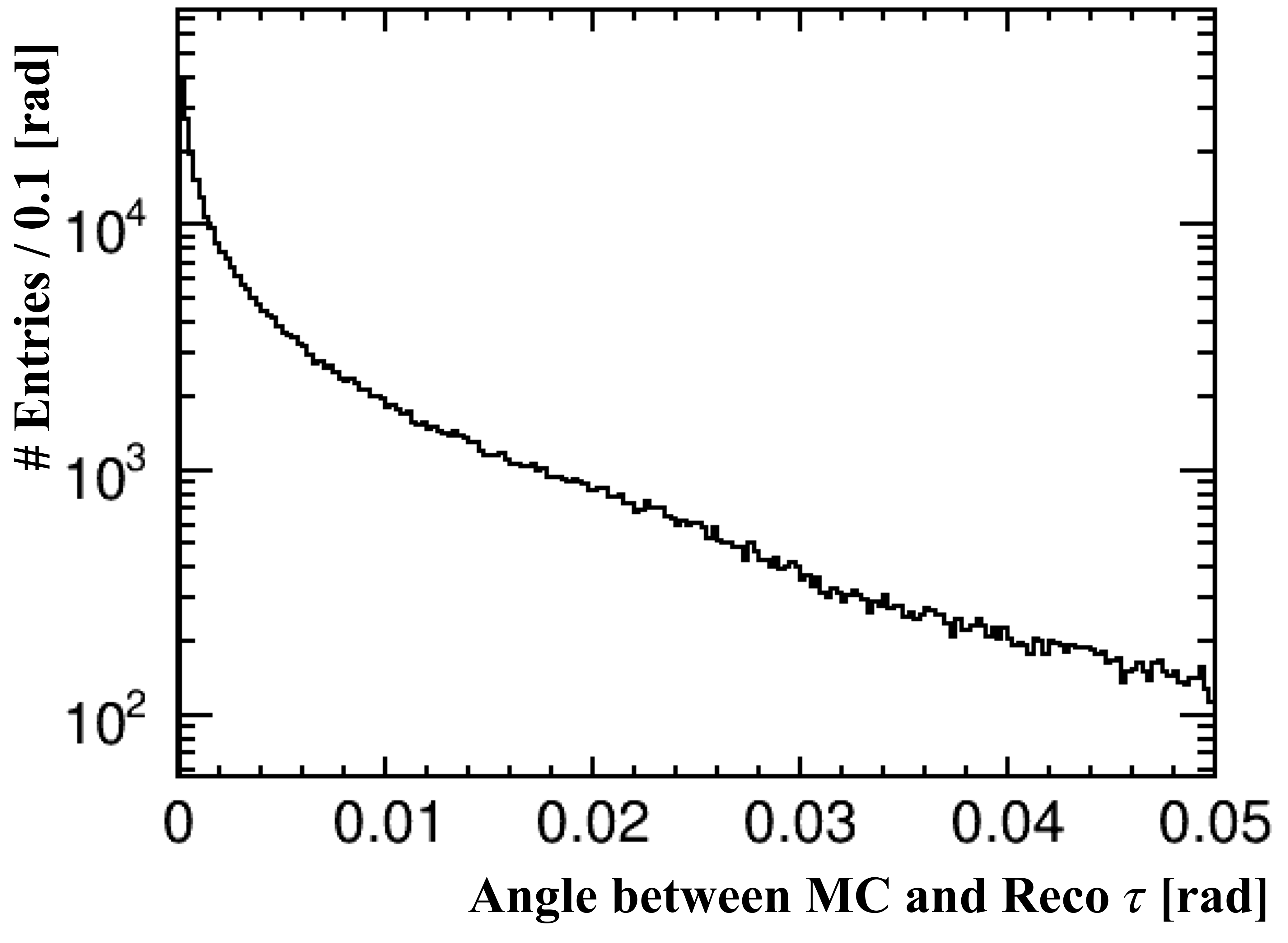}
  \caption{Top: The number of solutions, in all \eett\ events (left) and those with small ($<5$~GeV) ISR transverse momentum (right).
    Bottom: The angle between the true and reconstructed tau directions.
  }
  \label{nsol}
\end{figure}

The number of identified solutions in \eett\ events is shown in Fig.~\ref{nsol}, for all events, and for those in which ISR carries less
than 5~GeV of transverse momentum. Half of the method's inefficiency is due to cases in which there is large missing transverse momentum;
once these events are removed, the average efficiency to identify at least one solution is around 75\%.
The angle between the reconstructed and true tau directions is shown in the same figure: 
the reconstructed direction is typically within a few mrad of the true direction.

\begin{figure}[htb]
  \centering
  \includegraphics[width=0.4\textwidth]{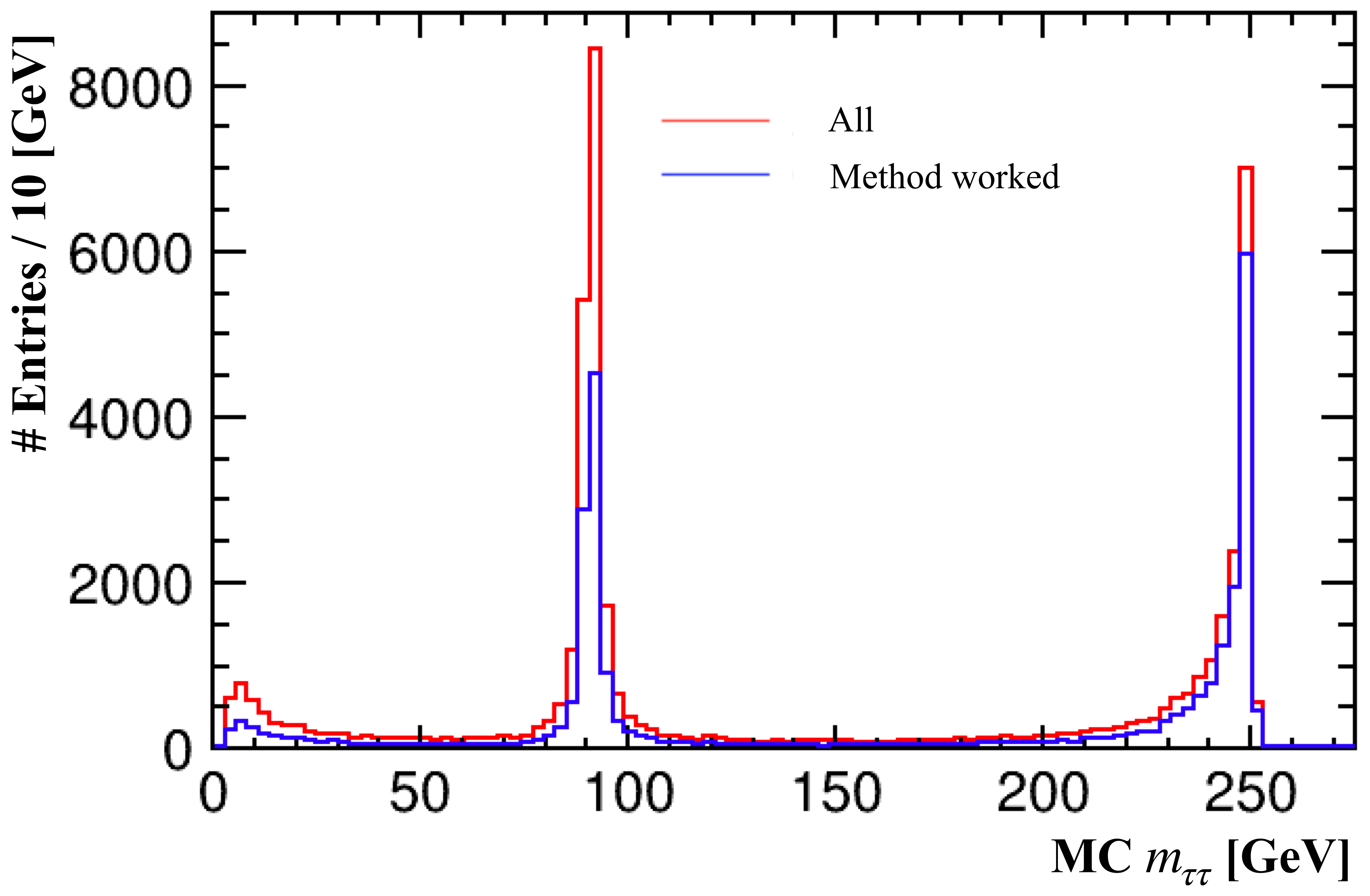} \\
  \includegraphics[width=0.4\textwidth]{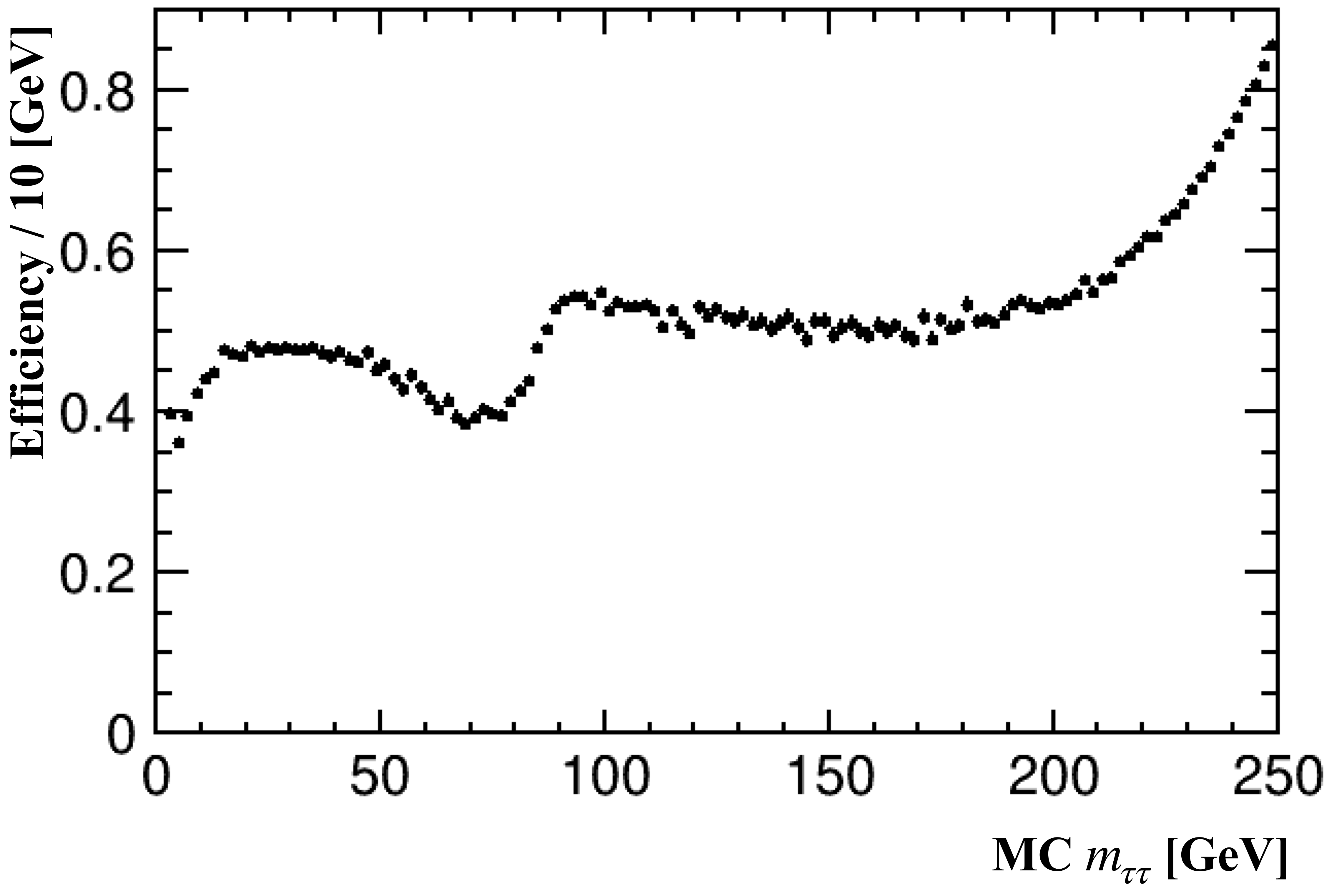}
  \includegraphics[width=0.4\textwidth]{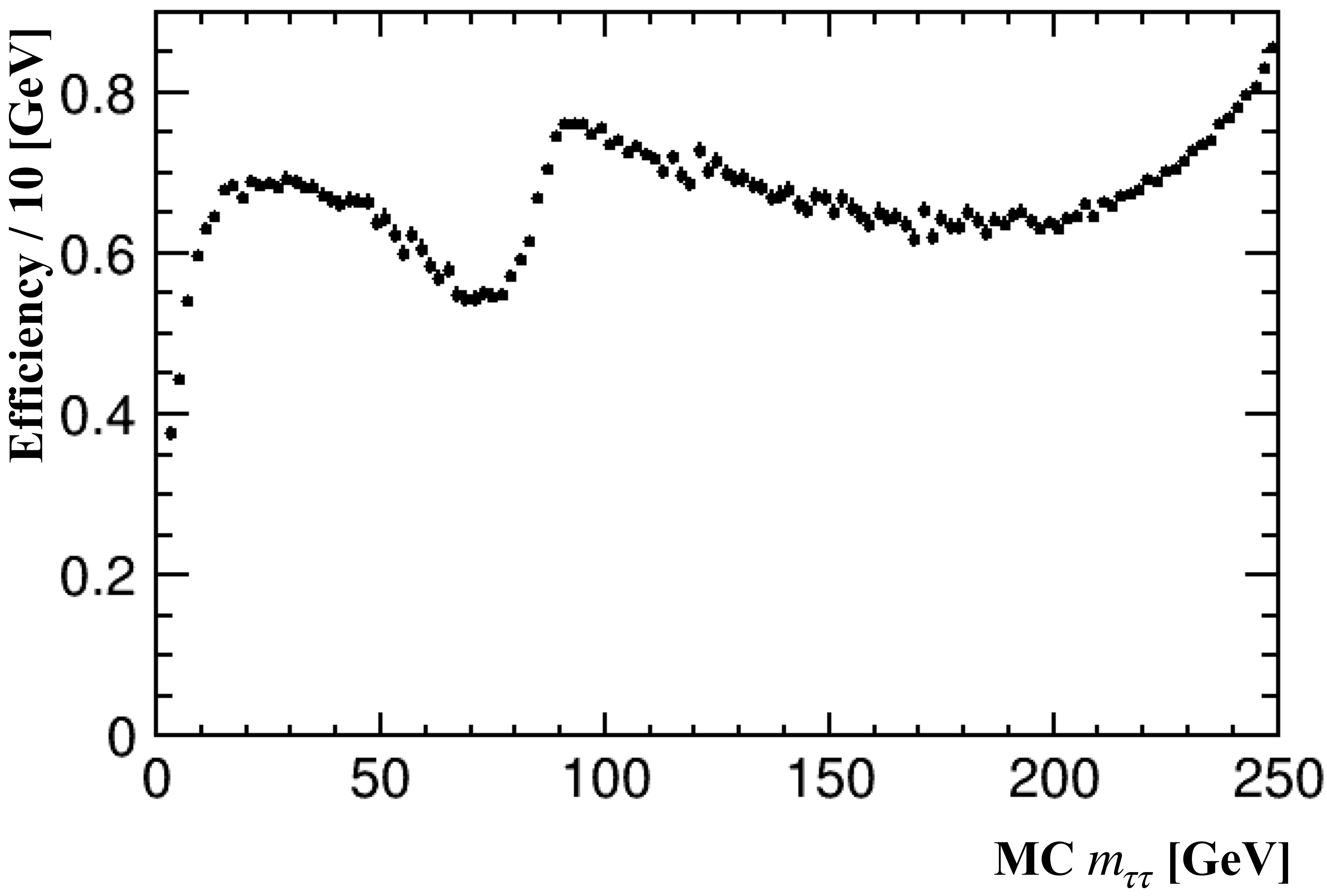}
  \caption{Upper: distribution of \eett\ events in \mtt, and distribution of events in which at least one ``impact parameter method'' solution was found.
Lower: \mtt\ dependence of the efficiency of the method in all events (left), and in events with small ISR transverse momentum (right).}
  \label{ISReff}
\end{figure}

Figure~\ref{ISReff} shows the distribution in \mtt\ of all events and of those in which at least one solution is found.
It also shows their ratio -- the method's efficiency -- for the entire sample, and when considering only events whose ISR
has small transverse momentum $<5~$GeV: such events can probably be experimentally identified by a search for visible photons far from the tau jets. 
The efficiency in this subsample is around 80\% at high \mtt, and remains in the range 60-70\% down to very small \mtt$\sim 10~$GeV.
There is an interesting structure around the $Z$ peak which is not yet fully understood.

\begin{figure}[htb]
  \centering
  \includegraphics[width=0.4\textwidth]{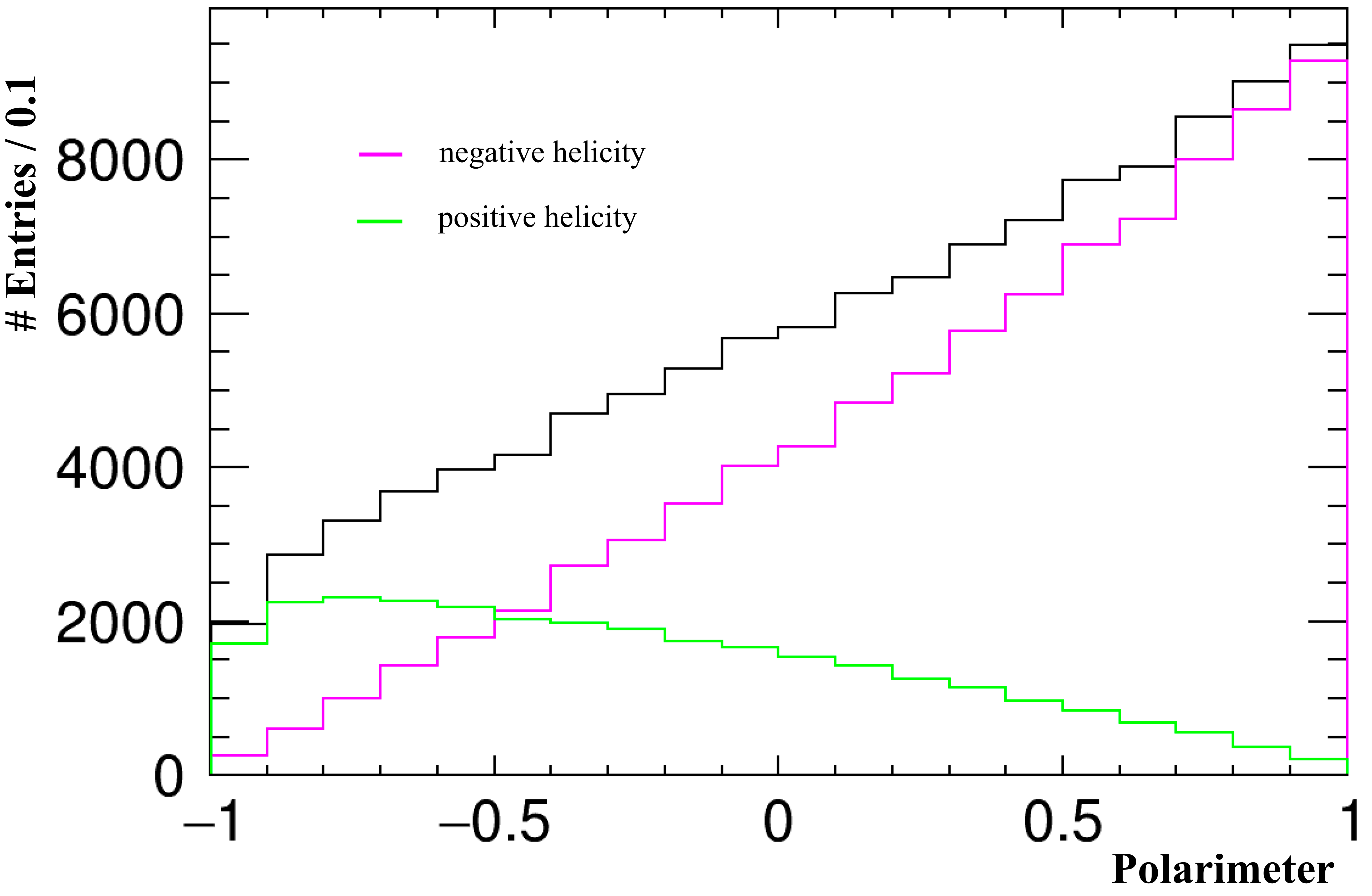}
  \includegraphics[width=0.4\textwidth]{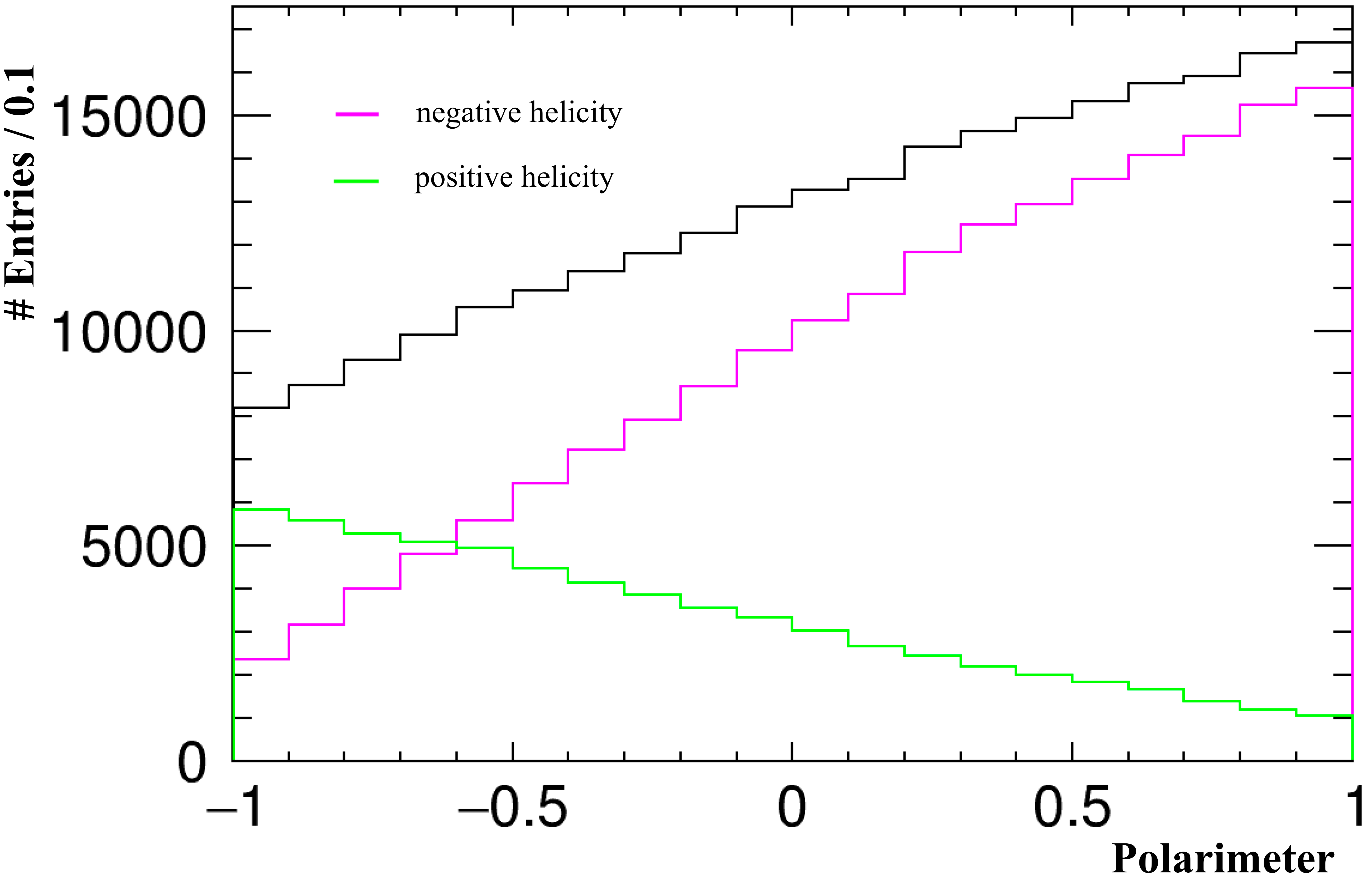} \\
  \includegraphics[width=0.4\textwidth]{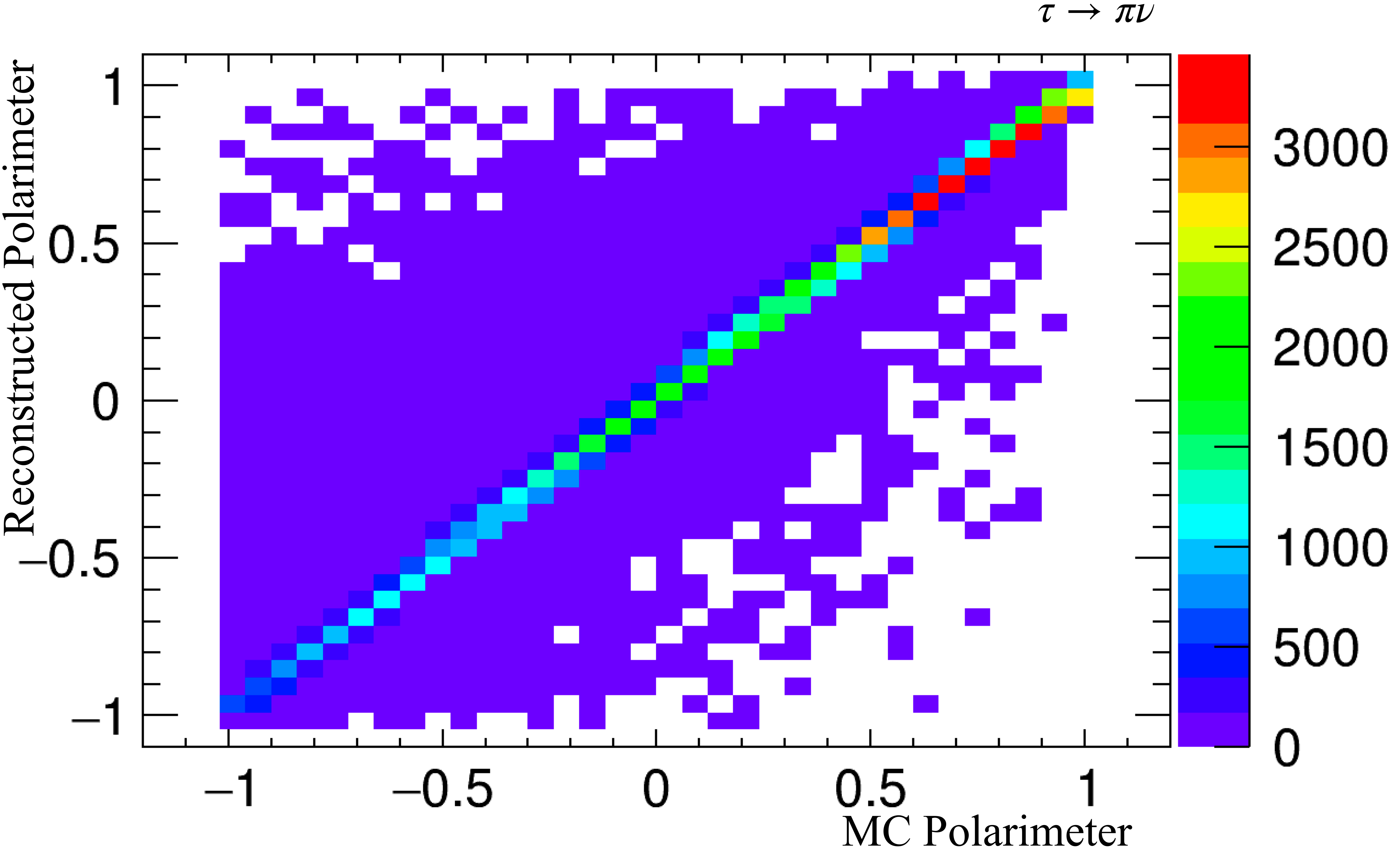}
  \includegraphics[width=0.4\textwidth]{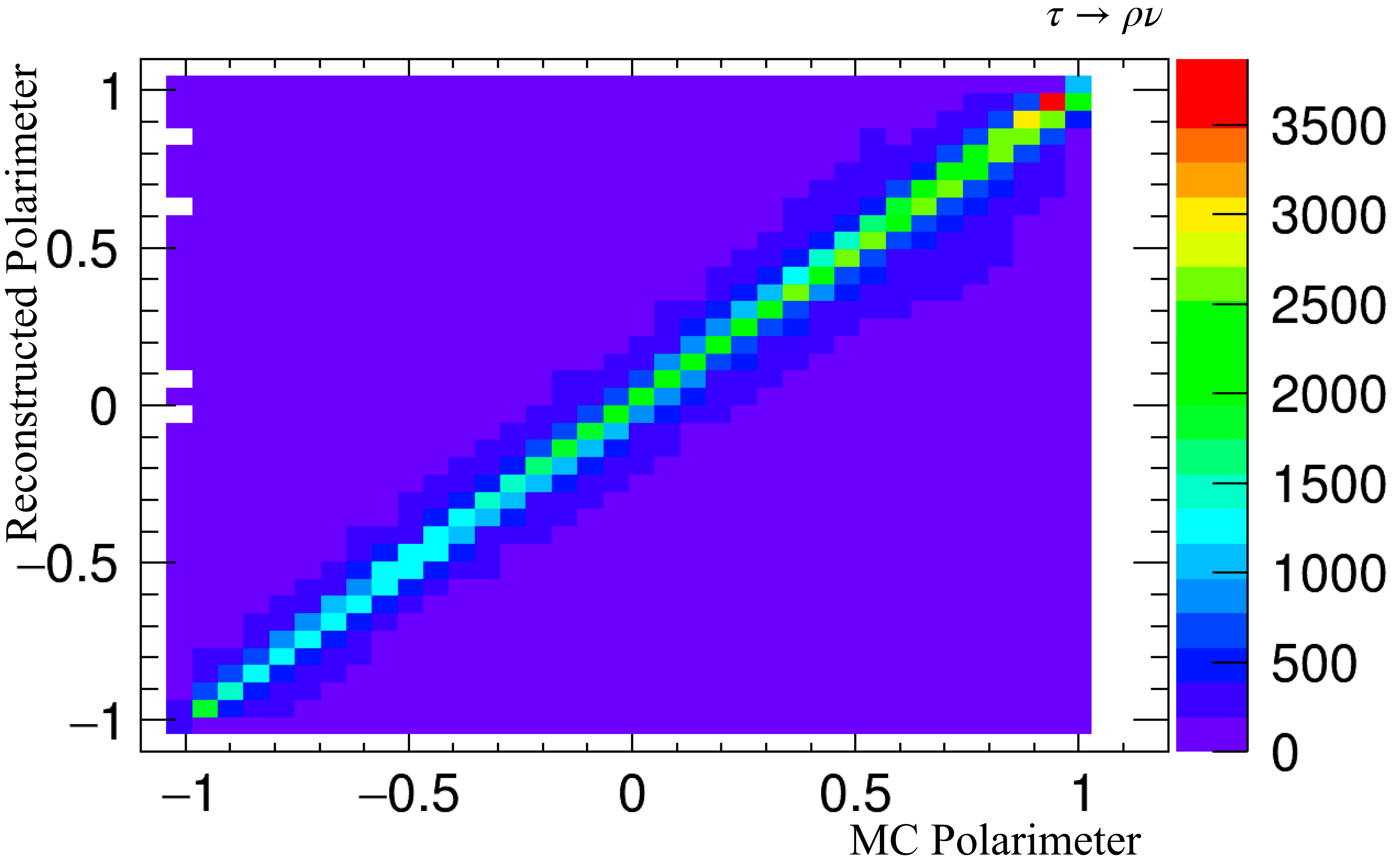}
  \caption{
    Top: Polarimeters for \tpn\ and \trn\ calculated using solutions found using the impact parameter method.
    Bottom: comparison between the reconstructed and true polarimeters.
}
  \label{imppol}
\end{figure}

Once the tau momenta have been reconstructed, polarimeter information can be calculated using the ``optimal'' forms 
with maximal sensitivity.
Figure~\ref{imppol} shows the distributions of polarimeters for taus of positive and negative helicity,
calculated using the solutions found by the impact parameter method.
A clear difference between taus of positive and negative helicity is seen.
The figure also shows the correlation between the true and reconstructed polarimeters,
which shows the rather good reconstruction quality for the majority of events, with only
a rather small probability of mis-reconstruction by a large amount.

These studies will be continued in the remaining period of the Snowmass study. We will estimate the
precision with which the tau polarisation can be measured at ILC-250, comparing the different reconstruction
techniques. At present, we ignore the effects of detector resolution: these will also be included in a future
update of this work.

\section{Conclusion}

We first reported on our previous full-simulation study which showed that the tau polarization can be measured to better than 1\% at ILC-500.

We further studied the reconstruction of tau polarisation at ILC-250, using particle momenta at MC level.
Several techniques to fully reconstruct \eett\ events were investigated, however they fail to accurately
reconstruct events with large energy loss through ISR.
We developed a new reconstruction method which makes use of charged particle impact parameter information,
as provided by the precise vertex detectors planned for ILC experiments.
In a study at the MC level, this method seems to provide efficient and accurate reconstruction of tau leptons,
allowing polarization information to be efficiently extracted.

In future work, we will investigate the effect of full detector simulation and reconstruction, 
quantify the precision with which the tau polarisation can be measured at ILC-250 at $\mtt \sim 250$ and $\sim m_Z$,
and compare the proposed reconstruction methods.

\bibliographystyle{unsrt}

\end{document}